\documentclass[twocolumn,showpacs,preprintnumbers,amsmath,amssymb]{revtex4}
\usepackage{graphicx}
\usepackage{dcolumn}
\usepackage{bm}
\begin{document}
\preprint{APS/123-QED}

\title{Sound beam through a circular aperture \\ and the far-field nonparaxial regime}

\author{Manas Kumar Roy}
\affiliation{S.N.Bose National Center for Basic Sciences\\
JD Block, Sector III, Saltlake, Kolkata-98, India\\
and Saha Institute of Nuclear Physics, 1/AF Bidhannagar, Kolkata-64, India}
\date{\today}

\begin{abstract}
Propagation of  sound beam diffracted from a circular aperture in far-field region has been studied in this paper by the method of angular  spectrum representation and stationary phase method. This nonparaxial theory is useful when beam angle is not very small and the wavelength  $\lambda $ is comparable  to aperture diameter $a$ unlike the situation in paraxial approximation. Here we have studied two cases, one for a Gaussian source and other for a plane piston source.
\end{abstract}

\maketitle

\section{\label{sec:level1}Introduction}
Pressure field radiation and  diffraction  from a source with a  circular aperture and its propagation through elastic medium is   quite well known and a well studied subject in acoustics\cite{blackstock},\cite{sample},\cite{hamilton}. Paraxial approximation has been  successfully applied {\cite{sample}} to study such phenomena. This approximation neglects  $\frac{\partial^2P}{\partial z^2}$ term (the second derivative of pressure  with respect to propagation axial co-ordinate) in the Helmholtz equation, as the term is of higher order in beam angle $\theta$ which is very small ($\sin\theta\approx\theta$).  But when $\theta$ is not so small, paraxial approximation is no longer valid. Some good works has already been done in optics in such nonparaxial correction \cite{duan},\cite{taka},\cite{fu}. But, in acoustics, such work is still missing. This paper concentrates on the propagation of the acoustic beam transmitted from a  monochromatic source from a circular aperture  in  nonparaxial  far-field zone, through a comparatively general approach employing the angular spectrum distribution and  the stationary phase method. Gaussian and plane piston  are taken as the  initial source. The paper begins with a discussion of an analytic nonparaxial formalism with analytical results. This is followed by a presentation of numerical results. Discussions with   conclusions are the final feature of this paper.
\section{Nonparaxial Theory}
\subsection{Helmholtz equation and angular spectrum representation} 
The governing equation for spatial part of pressure field is given by the Helmholtz equation
\begin{equation}
\lbrack\nabla^2 + k^2\rbrack P(x,y,z)=0
\end{equation}
In any plane $z=constant$, let us assume the field can be represented by a Fourier integral as
\begin{equation}
P(x,y,z)=\int_{-\infty}^{\infty}\int_{-\infty}^{\infty}\mathcal{P}(u,v;z)e^{ik(ux+vy)}dudv
\end{equation}
which when substituted in Eq.1 yields 
\begin{eqnarray}
\int_{-\infty}^{\infty}\int_{-\infty}^{\infty}\lbrack\nabla^2 + k^2\rbrack\mathcal{P}(u,v;z)e^{ik(ux+vy)}dudv=0
\end{eqnarray}
and then differentiating under integral sign, we find that
\begin{eqnarray}
\int_{-\infty}^{\infty}\int_{-\infty}^{\infty}\left[{(-u^2-v^2+k^2)\mathcal{P}(u,v;z)+\frac{\partial^2\mathcal{P}(u,v;z)}{\partial z^2}}\right]\nonumber\\
\times e^{ik(ux+vy)}dudv=0
\end{eqnarray}
For any values of $x$ and $y$, the term under bracket will go to zero. Hence the function $\mathcal{P}(u,v;z)$ will satisfy the following differential equation
\begin{equation}
\frac{\partial^2\mathcal{P}(u,v;z)}{\partial z^2} + w^2\mathcal{P}(u,v;z)=0
\end{equation}
Where $w^2=1-u^2-v^2$. When $u^2+v^2\le 1$, $w=+\sqrt{(1-u^2-v^2)}$. Again, when $u^2+v^2 > 1$, then $w=+i\sqrt{(u^2+v^2 -1)}$.\\
The general solution of the partial differential equation (5) is
\begin{equation}
\mathcal{P}(u,v;z)=A(u,v)e^{ikwz}+B(u,v)e^{-ikwz},
\end{equation}
where $A(u,v)$ and $B(u,v)$ are arbitrary functions of $u$ and $v$. On substitution of equation (2), we obtained
\begin{eqnarray}
P(x,y,z)&=&\int_{-\infty}^{\infty}\int_{-\infty}^{\infty}A(u,v)e^{ik(ux+vy+wz)}dudv+\nonumber\\
&&\int_{-\infty}^{\infty}\int_{-\infty}^{\infty}B(u,v)e^{-ik(ux+vy+wz)} dudv
\end{eqnarray}
If we don't consider reflection mode, then $B(u,v)=0$ (soft baffle case). Otherwise for rigid baffle, the pressure will be doubled due to reflection. So
\begin{equation}
P(x,y,z)=\int_{-\infty}^{\infty}\int_{-\infty}^{\infty}A(u,v)e^{ik(ux+vy+wz)}dudv,
\end{equation}
here $A(u,v)$ is defined as:
\begin{equation}
A(u,v)=\left(\frac{k}{2\pi}\right)^2\int_{-\infty}^{\infty}\int_{-\infty}^{\infty}P(x,y,0)e^{-ik(ux+vy)}dxdy
\end{equation}
Let, the source function be defined as
\begin{equation}
P(x,y,0)=P_0(x,y)\mathcal{T}(x,y)
\end{equation}
where $\mathcal{T}(x,y)$ is the window function for circular aperture defined as
\begin{eqnarray}
\mathcal{T}(x,y)=\left \{ \begin{array}{ll}
1 & \mbox{if $x^2+y^2\le a^2$} \\
0 & \mbox{otherwise}
\end{array}
\right .
\end{eqnarray}
Therefore $A(u,v)$ can be determined by
\begin{eqnarray}
A(u,v)=\left(\frac{k}{2\pi}\right)^2 \iint\limits_{x^2+y^2\le a^2}P_0(x,y)e^{-ik(ux+vy)}dxdy
\end{eqnarray}
Transforming from Cartesian coordinate to cylindrical coordinate system i.e. $x=\rho cos\theta$ and $y=\rho sin\theta$, and assuming the source is axially symmetric (Fig.1), the integral becomes
\begin{equation}
A(u,v)=\left(\frac{k}{2\pi}\right)^2\int_0^a\int_0^{2\pi}P_0(\rho,0)e^{-ik\rho(ucos\theta+vsin\theta)}\rho d\rho d\theta
\end{equation}
using the identity
\begin{equation}
\int_0^{2\pi}e^{-ikR\rho cos(\psi-\xi)}d\psi=2\pi J_0(kR\rho)
\end{equation}
\begin{eqnarray}
A(u,v)=\frac{k^2}{2\pi}\int_0^a P_0(\rho,0)  J_0(k\rho\sqrt{u^2+v^2})\rho d\rho
\end{eqnarray}
Thus $P(x,y,z)$ is
\begin{eqnarray}
P(x,y,z)&=&\frac{k^2}{2\pi}{\int_0^a}\iint\limits_{u^2+v^2\le a^2}P_0(\rho,0) J_0(k\rho\sqrt{u^2+v^2})\nonumber\\
&&\times e^{ik(ux+vy+wz)}\rho d\rho dudv
\end{eqnarray}
\subsection{Method of stationary phase}
The integral over u and v on the  right hand side of Eq.(16)
\begin{equation}
I={\int\int}_{u^2+v^2\le a^2} J_0(k\rho\sqrt{u^2+v^2})e^{ik(ux+vy+wz)}dudv
\end{equation}
may be approximated by  by method of stationary phase. Here $w=+(1-u^2-v^2)^{\frac{1}{2}}$ as $u^2+v^2\le1$, $k=\frac{\omega}{c}$. As the  evanescent or  exponentially decaying wave does not give any significant contribution in far field zone, we have avoided another condition $m=i(u^2+v^2-1)$ in our problem. Now let us specify some other variables $s_x=x/r$,  $s_y=y/r$,  $s_z=z/r$,
\begin{eqnarray}
I&=&\int\int_{u^2+v^2\le a^2}J_0(k\rho\sqrt{u^2+v^2})\nonumber\\
&&\times e^{ikr(us_x+vs_y+ws_z)}dudv,\\
&=&\int\int_{u^2+v^2\le a^2}a(u,v)e^{ikrg(u,v;s_x,s_y)}dudv
\end{eqnarray}
where $r=\sqrt{x^2+y^2+z^2}$, $a(u,v)=J_0(k\rho\sqrt{u^2+v^2})$ and $g(u,v;s_x,s_y)=s_xu+s_yv+s_zw$. The critical stationary points  for this integral will be $u_1=s_x$, $v_1=s_y$ and $w_1=s_z$. Hence in this approxination one obtains\cite{mandel}
\begin{eqnarray}
I&\sim& -\frac{2\pi i}{kr\sqrt{|\Delta|}}a(u_1,v_1)e^{ikrg(u_1,v_1;s_x,s_y)}\nonumber\\
&\sim& -\frac{2\pi i}{k}\left(\frac{z}{r}\right)J_0(k\rho\frac{\sqrt{x^2+y^2}}{r})\frac{e^{ikr}}{r},
\end{eqnarray}
where 
\begin{equation}
\Delta=(g_{uu}g_{vv}-g^2_{uv})_{u_1,v_1}
\end{equation}
$g_u=\frac{\partial g}{\partial u}$, $g_v=\frac{\partial g}{\partial v}$,  $g_{uu}=\frac{\partial^2g}{\partial u^2}$ etc.
Therefore equation(16) becomes
\begin{equation}
P(x,y,z)= \frac{k}{i}\left(\frac{z}{r}\right)\frac{e^{ikr}}{r}\int_0^a P_0(\rho,0)J_0(k\rho\frac{\sqrt{x^2+y^2}}{r})\rho d\rho,
\end{equation}
\section{Pressure fields for different axissymmetric sources}
\subsection{Gaussian source}
Sources with Gaussian amplitude is often been used because of  its functional form and invariant nature in different integral transforms. Let, the source function be\begin{figure}[h]
\includegraphics[height=60mm]{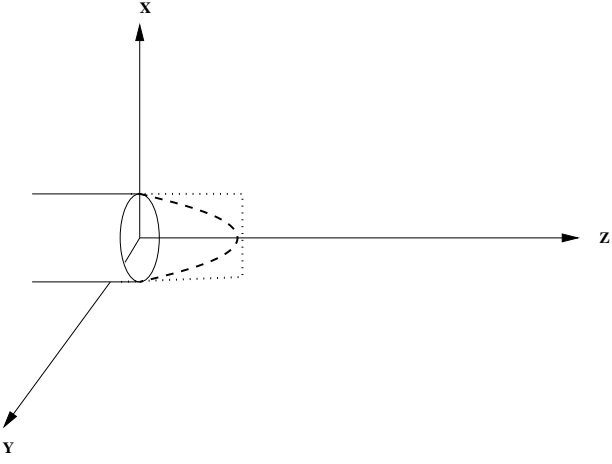}
\caption{\label{fig:epsart}Schematic diagram of the geometry of circularly apertured beam at $Z=0$ in Cartesian co-ordinate system. `- - -' line for Gaussian source, `....' line for piston source.}
\end{figure}
\begin{equation}
P_0(\rho,0)=P_0e^{-\frac{\rho^2}{a^2}}
\end{equation}
where $P_0=$ peak amplitude of the source. So, Eq.(22) becomes
\begin{equation}
P(x,y,z)= \frac{P_0 k}{i}\left(\frac{z}{r}\right)\frac{e^{ikr}}{r}\int_0^a e^{-\frac{\rho^2}{a^2}}J_0(k\rho\frac{\sqrt{x^2+y^2}}{r})\rho d\rho
\end{equation}
$J_0(x)$ can be expanded as -
\begin{equation}
J_0(x)=\sum_{m=0}^{\infty}(-1)^m\frac{\left(\frac{1}{4}x^2\right)^2}{(m!)^2}
\end{equation}
Thus, we can get{\cite{duan}}:
\begin{eqnarray}
P(x,y,z)=&&P_0\frac{i}{2k}\left(\frac{z}{r}\right)\left(\frac{e^{ikr}}{r}\right)\nonumber\\
&&\times\sum_{n=0}^{\infty}(-1)^n(ka)^{2n+2}\left(\frac{x^2+y^2}{4r^2}\right)^n\nonumber\\
&&\times\frac{\left[\gamma(1+n,1)-n!\right]}{(n!)^2}
\end{eqnarray}
where $\gamma(\alpha,\beta)$ is incomplete Gamma function defined as 
\begin{eqnarray}
\gamma(\alpha,\beta)&=&\int_0^\beta t^{\alpha-1}e^{-t}dt\nonumber\\
&=&\alpha^{-1}\beta^{\alpha}e^{-\beta}\phantom{x}_1F_1(1;1+\alpha;\beta)\nonumber
\end{eqnarray}
where $\phantom{x}_1F_1(1;1+\alpha;\beta)$ is the confluent hypergeometric function of the first kind. Eq.(26) is the basic analytical result for far-field  behavior of nonparaxial Gaussian beam with cirlcular aperture. Paraxial result can be obtained from this result by expanding $r$  and retaining  the first term in amplitude part, and retaining up-to the second term in phase part,
\begin{equation}
r\approx z+\frac{x^2+y^2}{2z}
\end{equation}
So that, Eq.(26) simplifies to
\begin{eqnarray}
P_{parax}(x,y,z)=&&P_0\frac{i}{2k}\frac{e^{ikz}}{z}e^{\left[\frac{ik}{2z}(x^2+y^2)\right]}\sum_{n=0}^{\infty}(-1)^n(ka)^{2n}\nonumber\\
&&\times\left(\frac{x^2+y^2}{4z^2}\right)^m\frac{[\gamma(1+n,1)-n!]}{(n!)^2}
\end{eqnarray}
Eq.(28) is the Fraunhofer diffraction formula for apertured Gaussian beam with a circular aperture  in paraxial regime.
 \begin{figure}[h]
\includegraphics[height=50mm]{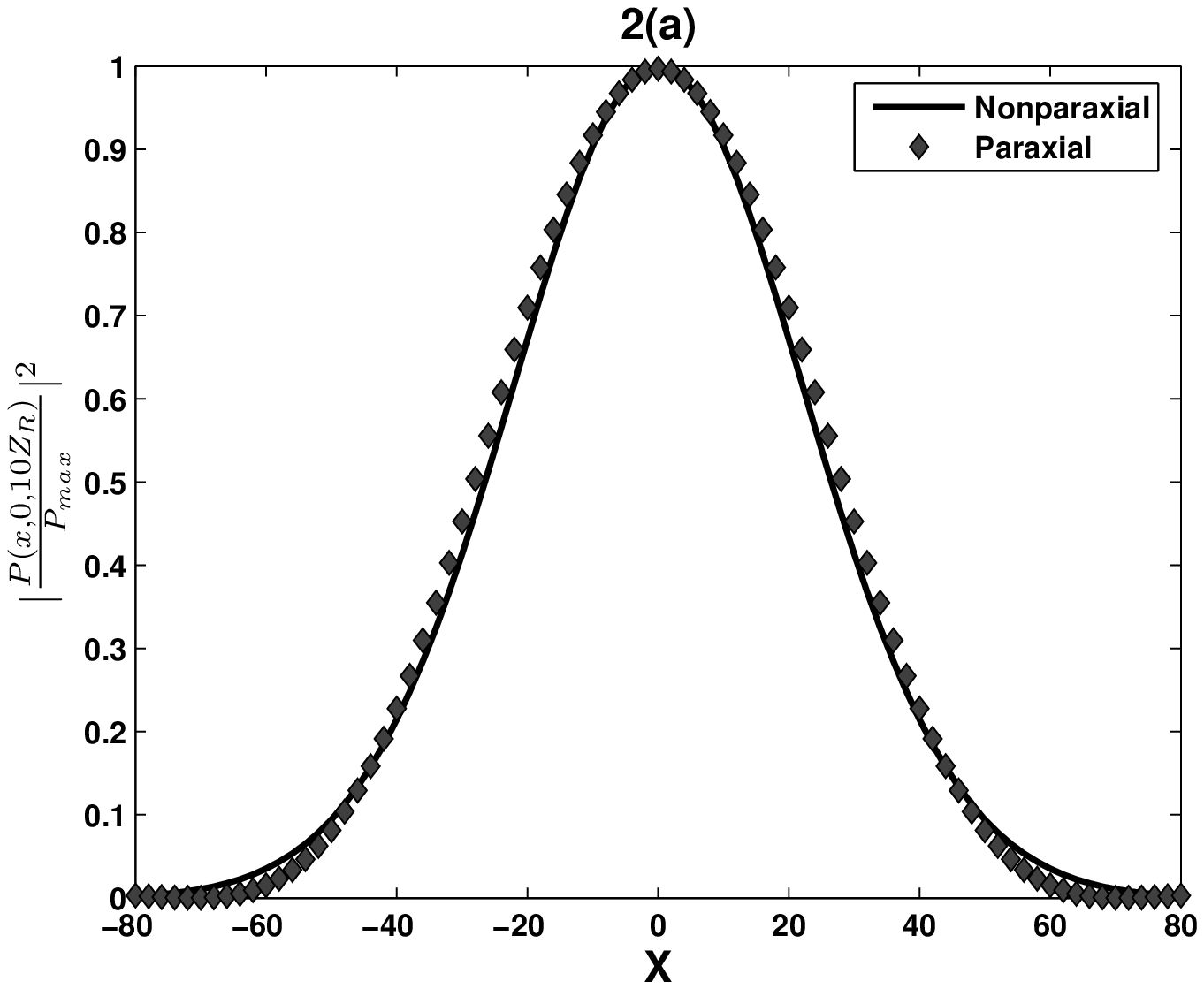}\\
\includegraphics[height=50mm]{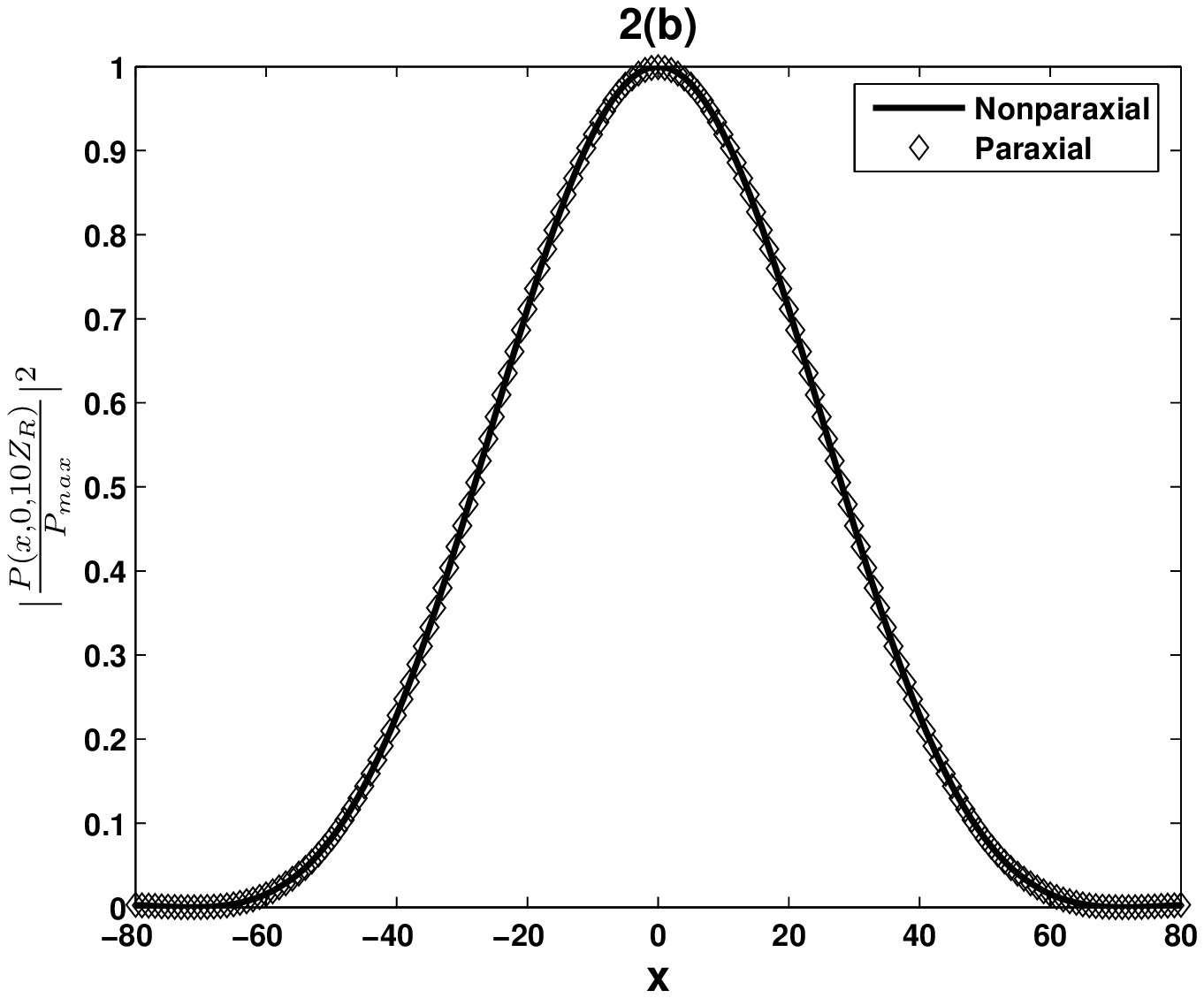}
\caption{\label{fig:epsart2}Distribution $|\frac{P(x,0,10Z_R)}{P_{max}(x,0,10Z_R)}|^2$ of a Gaussian sound beam diffracted from circular aperture of diameter $3.175 mm$.  2(a) frequency=$5\times 10^5$ Hz,  2(b) frequency=$5\times 10^6$ Hz}
\end{figure}

\subsection{Piston source}
In piston sources, the amplitude distribution can be described as\\
\begin{equation}
P(x,y,0)=P_0\mathcal{T}(x,y)
\end{equation}
Here the window function is same as Eq.(11). For this case, the resulting amplitude takes the integral form
\begin{eqnarray}
P(x,y,z)&=& -P_0{ik}\left(\frac{z}{r}\right)\frac{e^{ikr}}{r}\int_0^a J_0(k\rho\frac{\sqrt{x^2+y^2}}{r})\rho d\rho\nonumber\\
&=&-iP_0ka^2\left(\frac{z}{r}\right)\frac{e^{ikr}}{r}\left(\frac{r}{ak\sqrt{x^2+y^2}}\right)\nonumber\\
&&\times J_{1}\left(\frac{ak\sqrt{x^2+y^2}}{r}\right)
\end{eqnarray}
This is  another result derived in this paper for pressure field of nonparaxial circular apertured piston source in far-field region. Again, if we expand $r$ in series as given in Eq.(27), we can arrive at paraxially approximated result
\begin{equation}
P_{parax}=-iP_0ka^2\left(\frac{e^{ikz}}{z}\right)e^{\left[\frac{ik(x^2+y^2)}{2z}\right]}\left(\frac{J_1\left(\frac{ak\sqrt{x^2+y^2}}{z}\right)}{\left(\frac{ak\sqrt{x^2+y^2}}{z}\right)}\right)
\end{equation}
\begin{figure}
\includegraphics[height=50mm]{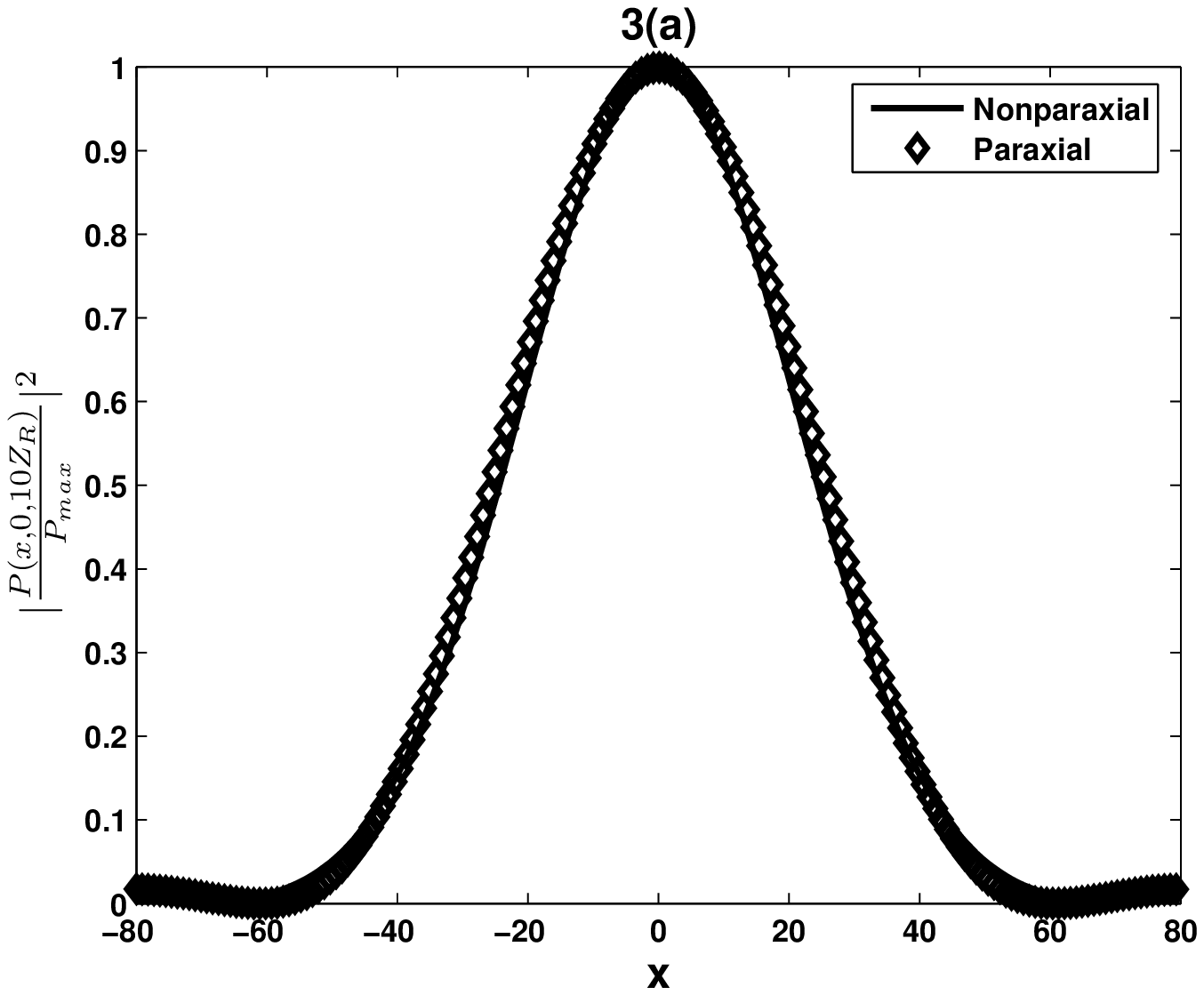}\\
\includegraphics[height=50mm]{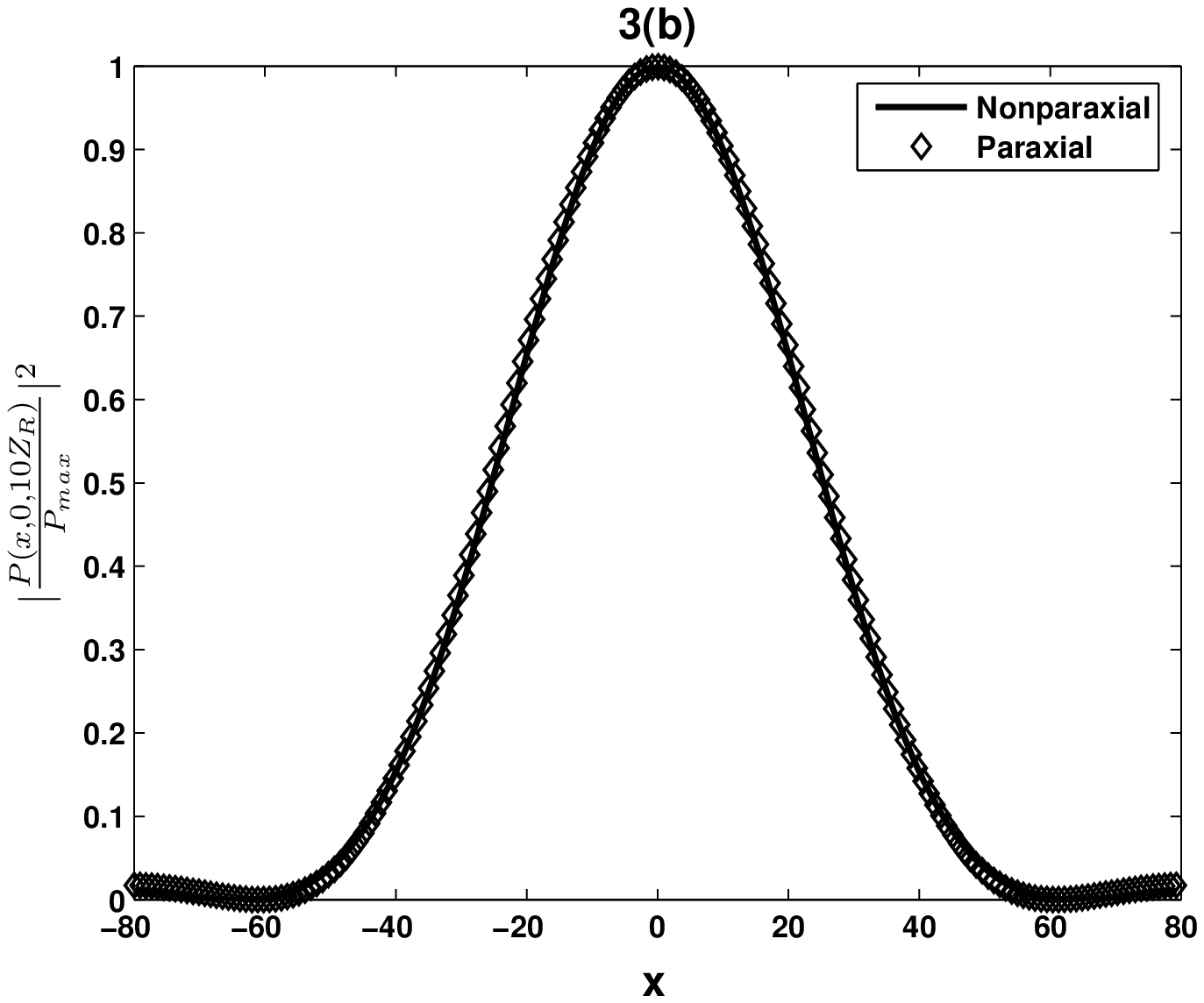}
\caption{\label{fig:apsart3}Distribution $|\frac{P(x,0,10Z_R)}{P_{max}(x,0,10Z_R)}|^2$ of a plane piston sound beam diffracted from circular aperture of diameter $3.175 mm$. 3(a) frequency=$5\times 10^5$ Hz,  3(b) frequency=$5\times 10^6$ Hz}
\end{figure}
Matlab 7.2  co
\section{Numerical Results}
des has been written to  solve Eq.(26), Eq.(28), Eq.(30) and Eq.(31). We have taken two different source frequencies, these are  $f_1=5\times10^5Hz$ and $f_2=5\times10^6Hz$. Aperture diameter has been taken as 3.175mm. Velocity of sound $c$ has been taken as 1500 m/sec.  The axial distances that has been taken to  compute the field is $10Z_R$, where $Z_R$ is Rayleigh distance defined as $Z_R=\frac{\pi a^2}{\lambda}$. The  axial distances are   $105.5641mm$ and $1055.6mm$ for $f_1$ and $f_2$ respectively.  Results has been shown in Fig.2 and Fig.3.
\section{Discussion}
An analytical method has been proposed to investigate sound beams beyond paraxial regime.  Angular spectrum representation and method of stationary phase are the two mathematical tools  used.  The analytical study and final results are for a general situation in far-field zone. Paraxial results can be obtained as the special cases of those results. Eq.(26) is the general nonparaxial solution for Gaussian source whereas Eq.(28) is its paraxial form. Similarly for the case of plane piston type source, Eq.(30) is the general nonparaxial solution and Eq.(31) is its paraxial form. Fig.(2) and Fig.(3) are the plots of the comparative numerical results for Gaussian and plane piston sources respectively. In Fig.2(a), nonparaxial and paraxial results differ. Although the maxima is same for nonparaxial and paraxial values, but the spread is  different. But as $\frac{a}{\lambda}$ increases [Fig.2(b)], this difference diminishes.  Similar result has been shown in Fig(3), i.e the results for plane piston source. We conclude that the analytical  results presented  here represents a  generalization of propagation of sound beam  diffracted from circular aperture in far-field zone. Not only generalization, but a significant correction over paraxial approximation too.
\section{Acknowledgment}
I would like to thank Prof.S.K.Sen, Mr.R.K.Saha, S.Karmakar and K.Ghosh of Saha Institute of Nuclear Physics, and Prof.S.K.Sharma, Prof.B.DuttaRoy of S.N.Bose National Center for Basic Sciences for this work. I wish to thank Council of Scientific \& Industrial Research (CSIR) for financial support. 


\begin{thebibliography}{1}
\bibitem{mandel} L. Mandel and E. Wolf, ``{\it{Optical coherence and quantum optics}}'', Cambridge University Press 1995
\bibitem{blackstock} D.T.Blackstock,  ``{\it{Fundamental of physical acoustics}}'', John Wiley \& Sons, 2000
\bibitem{duan} K. Duan, B. L$\ddot{u}$,  ``{\it{Nonparaxial analysis of farfield properties of Gaussian beams diffracted at a circular aperture}}'', Vol.11, No. 13, Optics Express (2003)
\bibitem{taka} T. Takenaka, M. Yokota, and O. Fukumitsu,``{\it{Propagation of light beams beyond the paraxial approximation}},'' J. Opt. Soc. Am. A 2, 826 (1985)
\bibitem{sample} John D. Sample,``{\it{Gaussian models for complex sound sources in the paraxial region}}'', J. Acoust. Soc. Am, Vol. 84, No. 6, December 1988
\bibitem{hamilton} M.F.Hamilton, V.A.Khokhlova and O.V.Rudenko, ``{\it{Analytical method for describing the paraxial region of finite amplitude sound beams}}'',
J.Acoust.Soc.Am, Vol.101, Issue 3, March 1997
\bibitem{born}M. Born and E. Wolf, ``{\it{Principles of Optics}}'' Pergamon, New York, 1980, 6th ed.
\bibitem{fu} Xiquan Fu, Hong Guo, Wei Hu and Song Yu, ``{\it{Spatial nonparaxial correction of the ultrashort pulsed beam propagation in free space}}'', Phys. Rev. E, Vol. 65, 056611 (2002)
\end{thebibliography}
\end{document}